\documentclass[12pt,preprint]{aastex}
\usepackage{natbib}
\bibpunct{(}{)}{;}{a}{}{,}

\shorttitle{A new case of super Li-rich K giant}
\shortauthors{Bharat $\&$ Reddy}

\begin{document}

\title{HD~77361: A new case of super Li-rich K giant with anomalous low $^{12}$C/$^{13}$C ratio }

\author{Y. Bharat Kumar and Bacham E. Reddy}
\affil{Indian Institute of Astrophysics, Bengaluru, India-560034}

\begin{abstract}

Results from high resolution spectroscopic analysis of HD~77361 are reported. 
The LTE analysis shows that HD~77361 is a K giant of atmospheric parameters: 
$T_{\rm eff}$ = 4580 $\pm$ 75 K, log ${\rm g}$ = 2.5 $\pm$ 0.1, 
and $\xi_{\rm t}$ = 1.40 $\pm$ 0.5 km s$^{-1}$. We found that the atmosphere of 
HD~77361 is highly enriched in Li with log $\epsilon$ (Li) = 3.82 $\pm$ 0.1. With this 
finding the total number of super Li-rich K giants (log $\epsilon$(Li) $\geq$ 3.3 ISM value) 
known till date reached six. Contrary to first dredge-up, extra-deep mixing and the 
associated cool bottom processing, and other recent predictions for K giants on the 
RGB luminosity bump phase, HD~77361 shows very low value of $^{12}$C/$^{13}$C = 4.3 $\pm$ 0.5
having, simultaneously, very large amount of Li. Also, HD~77361 is the only population~I 
low luminosity (log L/L$_{\odot}$ = 1.66 $\pm$ 0.1) 
low mass K giant (M = 1.5 $\pm$ 0.2M$_{\odot}$) among the known super Li-rich 
K giants  that has a very 
low  $^{12}$C/$^{13}$C  ratio. Results of HD~77361 further constrain our theoretical understanding
of Li enhancement in the atmospheres of RGB stars.
 
\end{abstract}


\section{Introduction}

Stellar evolutionary models predict significant depletion of Li in the atmospheres 
of evolved stars in which the convection layer reaches much deeper regions where 
temperatures are such that Li is destroyed. 
Standard first dredge-up stellar models 
predict Li reductions by 1-2 magnitudes in the low mass stars ($\leq$ 2.5M$\odot$) at the 
end of 1st dredge-up \citep{iben1967a,iben1967b} on the RGB phase from the main sequence maximum value of  
about log $\epsilon$(Li) =3.3 (e.g.,\citealt{lambert2004}). The maximum amount of Li one would expect, 
depending on mass, metallicity and the amount of Li on the main sequence, for a low mass K giant 
is  log $\epsilon$(Li) $<$ 1.5 dex \citep{iben1967a,iben1967b}. In fact, observations showed that most of 
the K giants of Pop I have much lower Li values  \citep{brown1989,mallik1999} 
compared to the predictions suggesting significant depletion of Li during its pre-main sequence 
and main sequence phases, and to some extent, due to non-standard mixing 
\citep{charbonnel1995}. The consensus among the investigators of Li in RGB stars is that a K giant 
with log $\epsilon$(Li) $\geq$ 1.5 can be termed as Li-rich. A giant with 
log $\epsilon$(Li) $\geq$ 3.3 (ISM value) may be termed as super Li-rich. 

Observations 
from different surveys reveal that Li-rich stars are very rare, and consists 
just 1$\%$ of  K giants \citep{brown1989} in the Galactic disk. Further, the super Li-rich 
K giants are much more rare, and as of now there are just six \citep{delareza1995,balachandran2000,drake2002,reddy2005}, 
including the one being studied here. To understand the distribution of Li-rich stars in 
terms of their mass and luminosity, \citet{charbonnel2000} located all the known 
Li-rich stars on the Hertzsprung-Russell diagram, thanks to the accurate parallaxes 
from $Hipparcos$ mission \citep{perryman1997}. The study revealed that the
(M/M$_{\odot}$ $\leq$ 2.0) super Li-rich (log $\epsilon$(Li) $\ge$ 3.3 dex) K giants are 
confined to a small region on the HR diagram, the so called luminosity 
bump: log L/L$_{\odot}$ = 1.45 - 1.9; log T$_{\rm eff}$ = 4450 - 4600 K) \citep{girardi2000}. 
Also, these stars are found to have lower $^{12}$C/$^{13}$C ratios than  
the expected from the standard mixing theory.
It is not well understood how the K giants with very deep convective envelope and efficient 
mixing, as suggested from the measured low values of $^{12}$C/$^{13}$C, possessed such 
high amounts of Li; almost 10 times larger than the ISM value. 
In the literature, a different class of super Li-rich stars were reported. They are either massive ($\geq$ 4M$_{\odot}$) weak G- band
stars \citep{lambert1984} or AGB stars \citep{smith1989} or low mass
metal-poor giants \citep{kraft1999,monaco2008}.

Li excess in K giants is one of the most puzzling stellar astrophysical problems to 
which, in the last two decades, great deal of research has been devoted. 
The high values of Li in the super Li-rich K giants are thought to be due to the stellar 
nucleosynthesis and some kind of extra mixing \citep{denissenkov2000,palacios2001,denissenkov2004}. 
At the luminosity bump on the RGB, the hydrogen burning shell passes the $\mu$-barrier 
or the mean molecular weight discontinuity that was created during the 1st 
dredge-up. This allows  
mixing between the cool outer layers and the hotter inner 
regions.
This so called 
extra-mixing theory 
associated with cool 
bottom processing was invoked for Li enhancement.
For low mass stars of solar metallicity at the luminosity bump, theoretical 
models (e.g., \citealt{denissenkov2004}) are constructed to match the observed peak Li 
abundance of log $\epsilon$(Li) $\geq$ 3 and low $^{12}$C/$^{13}$C $\approx$ 15-28. 
The theoretical results are, in general, in agreement with the observed values of 
super Li-rich K giants (\citet{charbonnel2000} and references there in). 
As star evolves from the bump region and moves up towards the tip of RGB, 
Li drops sharply with the $^{12}$C/$^{13}$C ratio \citep{lambert1980}.
 
HD~77361 is a bright field star of mv = 6.21 with colour B-V = 1.13. 
It is classified as a K1 III CN star \citep{houk1982}. This is one of the candidate 
K giants in our sample of 1800 stars that are selected from $Hipparcos$ catalogue 
\citep{vanleeuwen2007} for a systematic search of Li-rich K giants. 
The Li-richness of the star is identified from the low resolution spectra and later confirmed 
by obtaining series of high resolution spectra. Full details of the survey and results are presented
elsewhere.
We note that for HD~77361, no previous spectroscopic analysis of either low- or 
high resolution spectra is available. Results of very high Li abundance and the anomalously 
very low values of $^{12}$C/$^{13}$C ratios are quite puzzling given the low luminosity 
of HD~77361 and its location at the RGB bump region. This finding may further challenge 
the theoretical understanding of stellar structure and mixing mechanism during the RGB phase.

\section{Observations}

High quality low resolution spectra were obtained using medium resolution Zeiss 
universal astro grating spectrograph (UAGS) and 1 K $\times$ 1 K TEK CCD mounted on 
the 1~m Carl Zeiss telescope at Vain Bappu observatory, Kavalur. Followed by the detection 
of very strong Li resonance line at 6707\AA\ in the low resolution (R= $\lambda/\delta\lambda$ = 6000) spectrum of HD~77361, 
high resolution spectra were collected from 2.3m Vainu Bappu Telescope (VBT)  
equipped with fiber fed coude echelle spectrograph \citep{rao2005} and 2 k $\times$ 4 K CCD. 
Final spectral resolution as measured from arc spectrum 
is R$\approx$60,000. Most of the spectra observed with VBT were under poor weather 
conditions, and hence relatively poor quality (S/N$\approx$100) given the brightness 
of the star (mv = 6.21). Also, one set of high quality (S/N$\approx$500) echelle spectrum 
with resolution R$\approx$55,000 was obtained from cross-dispersed echelle spectrometer 
mounted on Harlan J Smith 2.7m telescope at McDonald Observatory. Standard Echelle reduction
procedure was adopted using the $IRAF$ software. 

\section{Analysis and Results}

Local thermodynamic equilibrium  (LTE) stellar model atmospheres 
computed by \citet{kurucz1994}\footnote[1]{www.kurucz.harvard.edu} with convection option on, and the revised radiative 
transfer code MOOG originally written by \citet{sneden1973} were adopted for the analysis. 
Standard procedure of requiring the same abundances from the neutral lines of different 
low excitation potentials was used for $T_{\rm eff}$ determination. Forty  well 
separated and moderate strength (W$_{\lambda}$ $\leq$ 100m\AA) Fe\,{\sc i} and eight Fe\,{\sc ii} lines 
\citep{allende2002,reddy2003} with accurate $gf$-values were used in 
the analysis. Spectroscopically  derived  $T_{\rm eff}$ = 4580 $\pm$ 75 K was found to be 
in good agreement with the photometric values; 4550 K (B-V) and 4587 K (V-K), 
which were determined using the  \citet{alonso1999} calibrations. 
The accurate parallaxes combined with the evolutionary tracks \citep{girardi2000} were used 
to determine the value of surface gravity of log $g$ =2.5 $\pm$ 0.1. This is consistent with 
the log $g$ value obtained by forcing Fe\,{\sc i} and Fe\,{\sc ii} lines to yield the same abundance 
for a given $T_{\rm eff}$ but for varying log $g$ values. Value of microturbulence 
$\xi_{\rm t}$ = 1.4 $\pm$ 0.5 km s$^{-1}$ was extracted by requiring to have the same 
abundance for a set of Fe\,{\sc i} lines of different W$_{\lambda}$. 
Metallicity, [Fe/H] = $-$0.02 $\pm$ 0.1, was determined from the mean 
abundance of Fe\,{\sc i} and Fe\,{\sc ii} relative to the sun \citep{lodders2003}.
Radial velocity measurements made over a period of 
20 days do not suggest variation in the velocity. 
The mean radial velocity relative to helio centric velocity was found to 
be $R_{\rm v}$ = $-$23.20 $\pm$ 0.5 km s$^{-1}$.  

Abundances of Li, CNO elements, and $^{12}$C/$^{13}$C ratios were derived by matching 
the observed spectrum with that of the computed spectrum. The well tested line 
list \citep{reddy2002} in the region of Li resonance line 6707.8\AA\ was adopted. 
Hyperfine structure and the $gf$ values of $^{7}$Li components of Li line at 6707.8\AA\ 
were taken from \citet{hobbs1999}. Computed spectrum was convolved
with estimated values of microturbulance of 1.4 km s$^{-1}$, 
macroturbulance of 3 km s$^{-1}$, and rotational velocity of 3 km s$^{-1}$.
A value of log $\epsilon$(Li) = 3.96 dex was obtained 
from the resonance line (Fig~1) assuming pure $^{7}$Li abundance. The Li line at 6103.6\AA\, 
normally weak in the intermediate Li rich stars, is very strong (W$_{\lambda}$ = 151m\AA) 
in the spectrum of HD~77361. The compiled line list from Kurucz's line database was used for 
computing the spectrum at the vicinity of Li excited line at 6103.6\AA. For Li abundance 
of log $\epsilon$ (Li) = 3.67 dex, the synthetic spectrum matches well with the observed 
spectrum (Fig~1). Significant difference between the abundances of Li lines 
at 6707\AA\ and 6103\AA\ is due to severe non-LTE effects ($\Delta$(LTE-NLTE) = 0.16 dex for 6707 line
and $-$0.16 dex for 6103 line) acting in the opposite 
directions \citep{carlsson1994,lind2009}. The LTE abundances, and the abundances 
after correcting NLTE effects are given in Table~1. Corrected values of Li from the two lines 
agree very well (Table~1).  
Li abundance of 
log $\epsilon$(Li) = 3.82 for HD~77361 is the straight mean of Li abundances, 
after non-LTE corrections, from the two Li lines at 6103\AA\ and 6707\AA. 
Same analysis is duplicated for the comparison star HD~19745, and the results obtained 
are consistent with the previous studies \citep{reddy2005}. 

The $^{12}$C/$^{13}$C ratio is proved to be a very sensitive parameter of the 
stellar mixing process. 
Measuring $^{12}$C/$^{13}$C ratio along with 
the Li abundance has an obvious advantage in understanding the RGB mixing process. 
As shown in the Fig~2, the $^{13}$C$^{14}$N lines at 8004.6\AA\ together with a 
few $^{12}$C$^{14}$N lines in the same region were used for deducing $^{12}$C/$^{13}$C ratio. 
Abundance  of carbon isotope $^{12}$C was derived by matching the observed spectrum with that 
of the computed spectrum of C\,{\sc i} line at  5380\AA\ , and $C_{2}$ Swan system features at 5086\AA\ and 5135\AA\ .
The mean abundance of the three lines log $\epsilon$ = 8.45 $\pm$ 0.05 was used to determine 
the N abundance from CN lines in the 8003\AA\ region. Oxygen abundance was 
determined from the two forbidden lines at 6300\AA\ and 6363\AA. The line list and 
the procedure adopted is very similar to that was used in our earlier studies 
\citep{reddy2005}. Abundances of CNO and $^{12}$C/$^{13}$C ratio 
are given in Table~1. The quoted errors in the abundances are the quadratic sum of 
uncertainties due to the uncertainties in the model parameters.  

Accurate $Hipparcos$ parallaxes ($\pi$ = 9.25 $\pm$ 0.43 mas) are available for HD~77361 
\citep{vanleeuwen2007}. Mass (1.5~M$_{\odot}$) and luminosity (log L/L$_{\odot}$ =1.66) 
were obtained combining the parallax information 
with photometry and evolutionary tracks \citep{girardi2000} of solar 
metallicity. Four stars
in the table 
lack parallax information for which values of mass and luminosity were derived 
by combining spectroscopy and photometric information 
\citep{balachandran2000,drake2002,reddy2005}. HD~77361, 
along with the known super Li-rich stars, is shown on the HR diagram (Fig~3).  
All the six K giants occupy the region just around the RGB luminosity bump which 
is marked in the figure~3.

\section {Discussion} 

Results of five super Li-rich K giants are given along with HD~77361 
in Table~2. Values of other stars are taken from the literature. 
For the most Li-rich K giant HD~233517, unfortunately,  the $^{12}$C/$^{13}$C ratio is 
not available, probably, due to its very high rotation of $vsini$ = 17 km s$^{-1}$. 
Generally, the $^{13}$C$^{14}$N lines at 8004.6\AA, often used for the carbon isotopic ratio,  
is weak and gets smeared out in the spectra of high rotation stars. 

It is clear from the Table~2 that HD~77361 is a super Li-rich, and the 3rd most Li-rich K 
giant amongst six super Li-rich K giants known so far. The derived luminosity 
(log L/$L_{\odot}$ = 1.66) and $T_{\rm eff}$ = 4580 K places HD~77361 at the RGB luminosity bump. 
The most striking from the results shown in Table~2 is that the very low $^{12}$C/$^{13}$C 
ratio of HD~77361. The other super Li-rich stars have significantly higher values 
of $^{12}$C/$^{13}$C ratios. It is puzzling that HD~77361 has a $^{12}$C/$^{13}$C ratio 
that is close to the CN equilibrium value, but continues to have peak value of Li 
on its surface.
Another important point to be noticed is that the location of super Li-rich stars. 
All are very close to the expected luminosity bump region for a given stellar mass. 
Results of HD~77361 
strengthens the argument that the changes to the stellar structure of low mass 
stars at the bump region are the main cause for the sudden reduction in $^{12}$C/$^{13}$C 
ratio, and the enhancement of Li on the surface of RGB stars.

It is known that at the RGB bump the hydrogen burning shell moves past 
the so called $\mu$-barrier. In the absence 
of $\mu$-barrier, convective zone reaches much deeper level dredging up the hydrogen 
burning products to the surface altering abundances of Li, C, N, and $^{13}$C much more 
severely compared to the predictions of standard 1st dredge-up scenario. This is known 
as extra mixing \citep{sweigart1979,charbonnel1998} which is different 
from the ordinary convective mixing. The extra mixing is invoked to explain the observed, 
in general, low values of Li and $^{12}$C/$^{13}$C ratio at or above the RGB bump 
\citep{lambert1980,gratton2000}. The exact mechanism for the extra mixing is still an issue to 
be worked out. 

For the Population I stars of low mass (M/M$_{\odot}$ $\leq$ 2.5) theoretical models of extra mixing 
predict final $^{12}$C/$^{13}$C ratios as low as 10 when star evolves to the RGB tip. 
\citet{boothroyd1999} demonstrated that deep circulation mixing at the base of 
convective envelope and the associated cool bottom processing brings down the 1st dredge-up 
prediction of $^{12}$C/$^{13}$C$\approx$ 28 to the lowest final value of 10 closer to the RGB tip for a star 
of 1.5 M$\odot$ with solar metallicity. 
Similar results from the recent computations based on enhanced extra mixing 
triggered by external sources \citep{denissenkov2004} and mean 
molecular weight inversion or $\delta$-mixing \citep{eggleton2008} predicted final values 
of $^{12}$C/$^{13}$C $\approx$ 10-14 for low mass Pop I stars. 
However, in all the models, stars at the RGB bump region are expected to have $^{12}$C/$^{13}$C 
ratios at about 15 or more.

Earlier suggestions of planet or dwarf companion merger \citep{alexander1967,brown1989,siess1999} with 
the parent stars or addition of Li rich material from the nearby novae ejections 
\citep{gratton1989} cannot explain the localization of super Li-rich stars on the 
HR diagram. If the Li enhancement in the K giant is due to the external sources like 
planet engulfment or accretion of novae material, it could happen anywhere on the RGB phase. 
Absence of low mass super Li-rich giants below or above the RGB bump simply means 
that Li enhancement is related to changes in the internal stellar structure and the mixing 
process that are associated with the luminosity bump. Depletion of Li starts once again,
from the peak Li abundance at the RGB bump, 
as stars evolve towards the RGB tip. Observations show, in general, very low Li abundances 
(log $\epsilon$(Li) = $-$1.0 - 1.5 dex) in K giants between the bump and the RGB tip. 
As discussed by \cite{delareza1997} and calculated by \cite{denissenkov2004}
the super Li-rich phase on the RGB seems to be very short (a few million years), 
and all the low mass stars, perhaps, experience this phase. 

Another suggestion one can infer from the results shown in Table~2 is that the absence 
of correlation between Li enhancement and the $^{12}$C/$^{13}$C ratio. 
It seems Li enhancement in the K giants at the bump is independent of 
the $^{12}$C/$^{13}$C ratio, and hence longevity or the extent of the 
deep mixing process. 
\cite{sackmann1999} suggest that Li can be enhanced
on the surface as long as fast deep mixing is ensured, and in fact they have
predicted high values of Li towards the RGB tip (not at the RGB bump) 
where $^{12}$C/$^{13}$C values are low. On the other hand, \cite{palacios2001}
hypothesized that the surface $^{7}$Li enhancement is the precursor for the extra mixing.
Given the very low value of $^{12}$C/$^{13}$C ratio for the HD~77361, 
one may suggest that
erasing the $\mu$-barrier, and hence the free mixing 
of material between hydrogen burning shell and the bottom of convective outer layer seems 
to be providing conducive environment for the enhancement of surface Li abundance. 

It is understood that Li is produced in the inner layers of the stars through 
\cite{cameron1971} mechanism. 
Sustaining the Li production and dredging up 
the Li rich material to outer layers to the observed levels is still an unsolved problem. 
For safe transportation of $^{7}$Be produced in the inner layers via 
$^{3}He(\alpha,\gamma)^{7}$Be to cooler regions, so that $^{7}$Li could be produced 
and mixed up with the photosphere, a few possibilities were explored. 
\cite{palacios2001} suggested Li-flash scenario for high Li-abundance. 
In this scenario, $^{7}$Be diffuses outward where Li can be produced 
via $^{7}Be(e^{-},\nu)^{7}$Li and forms the so called Lithium burning shell. 
Energy release from the destruction of $^7$Li by proton 
as well as from production of $^{7}$Li  destabilize the Li burning shell 
leading to runway situation. This in turn leads to 
the enhanced meridional 
circulation
and hence the observed Li enhancements. 
In this scenario Li enhancement precedes the 
$^{12}$C/$^{13}$C  reduction. This explains the localization of super Li-rich
K giants at the bump and also the relatively high values of carbon isotopic
ratios \citep{charbonnel2000}. As K giants evolve away from the bump Li starts decreasing with the
decreasing $^{12}$C/$^{13}$C ratio. A totally different approach was suggested from the
computations by \cite{denissenkov2004}. 
An enhanced extra mixing triggered by spinning up of the K giants with the external
angular momentum can produce the required mixing coefficients. Source of the external
angular momentum could be either by the synchronization of K giant's spin with
the orbital motion of close binary or engulfment of massive planet. This mechanism
will enhance K giant's angular momentum by 10-fold, and explains high rotation velocity
observed in some of the Li-rich K giants. But,
as stated earlier, fails to
explain the presence of all the six super Li-rich  at the bump and the absence of super Li-rich
stars any where along the RGB path. 
In all of these models, Li peak was achieved when 
the $^{12}$C/$^{13}$C ratios are around 15 - 28 which are in general agreement 
with the five of the six super Li-rich stars shown in the Table~2. 

The exception being the results of HD~77361 presented in this paper.
It is the only bump K giant that has very 
low $^{12}$C/$^{13}$C ratio and still continuous to have peak Li abundance. This result 
may put further constraints on the RGB models. Peak Li abundance (log $\epsilon$ (Li) = 3.82 dex) 
on the surface and the lowest $^{12}$C/$^{13}$C = 4.3 of HD~77361 suggest that whether the removal of $\mu$-barrier
and hence the free deep mixing is essential for the Li enhancement. 
To understand the exact mechanism that is responsible for the observed Li enhancements
on the K giants, it is worth to make a systematic survey of  low mass (M $\leq$ 2.5 M$_{\odot}$ ) 
K giants all the way from the start of 1st dredge-up to the tip of the RGB.

\acknowledgments
We are grateful to David Lambert for obtaining the high quality spectrum for HD~77361. 
We also thank VBO observatory staff for their help during the observations.

\newpage

\begin{table*}
\begin{center}
\caption{ Abundances of Li and CNO elements for HD~77361 and HD~19745.
\label{tbl-1}}
\begin{tabular}{ccccrrrcr}
\tableline\tableline
Star     & [Fe/H]$^{a}$ & [C/Fe]$^{a}$ & [N/Fe]$^{a}$& [O/Fe]$^{a}$ & \multicolumn{2}{c}{log $\epsilon$(Li)$_{LTE}$ } & log $\epsilon$ (Li)$_{NLTE}$ &
$^{12}$C/$^{13}$C  \\
          &        &       &    &    & Li 6103\AA\  & Li 6707\AA\  & (mean)  \\
\tableline
HD~77361  & $-$0.02$\pm$0.1 & 0.01  & 0.076 & 0.039  & 3.67$\pm$0.10  &  3.96$\pm$0.14 & 3.82 &  4.3$\pm$0.5  \\
HD~19745  & $-$0.05$\pm$0.1& 0.00  & 0.047 & 0.017  & 3.91$\pm$0.12  & 3.69$\pm$0.27  & 3.77 & 15$\pm$2  \\
\tableline
\end{tabular}
\tablenotetext{a}{Adopted Solar values are recommended abundances from \cite{lodders2003}}
\end{center}
\end{table*}

\newpage

\begin{table*}
\begin{center}
\caption{ Values of stellar parameters (Metallicity, $T_{\rm eff}$, Mass, luminosity) and
the surface abundances values of log $\epsilon$(Li) and carbon isotopic ratios of HD~77361
are given along with the known super Li-rich K giants.
\label{tbl-2}}
\begin{tabular}{cccrrrrr}
\tableline\tableline
Star     & [Fe/H] & T$_{\rm eff}$
& M$\star$/M$_{\odot}$ & log  L/L$_{\odot}$ & log $\epsilon$(Li) & $^{12}$C/$^{13}$C & Ref.  \\
\tableline
HD~77361  & $-$0.02$\pm$0.1 & 4580$\pm$75 & 1.5$\pm$0.2  &  1.66$\pm$0.1 &  3.82$\pm$0.10 &  4.3$\pm$0.5   & This paper  \\
         &               &              &       &               &                &            &   \\
HD~233517 &  $-$0.37      & 4475$\pm$70 &   1.7$\pm$0.2  &  2.0$^{a}$          &  4.22$\pm$0.11 &  ...       & 1  \\
IRAS~13539-4153 & $-$0.13      & 4300$\pm$100 &  0.8$\pm$0.7  &  1.60$^{a}$         &  4.05$\pm$0.15 &  20        & 3  \\
HD~9746 &    $-$0.06      & 4400$\pm$100 &  1.92$\pm$0.3  &  2.02         &  3.75$\pm$0.16 &  28$\pm$4  & 1  \\
HD~19745 &  $-$0.05       & 4700$\pm$100 &  2.2$\pm$0.6  &  1.90$^{a}$         &  3.70$\pm$0.30 &  16$\pm$2  & 3  \\
IRAS~13313-5838 &  $-$0.09       & 4540$\pm$150 &  1.1  &  1.85$^{a}$  &  3.3$\pm$0.20 &  12$\pm$2  & 2  \\
\tableline
\end{tabular}
\tablenotetext{a}{Hipparcos astrometry is not available and values of luminosities and masses are derived from spectroscopy. }
\tablerefs{(1) \citealt{balachandran2000}. (2) \citealt{drake2002}. (3) \citealt{reddy2005}. }
\end{center}
\end{table*}

\newpage

\begin{figure}
\plotone{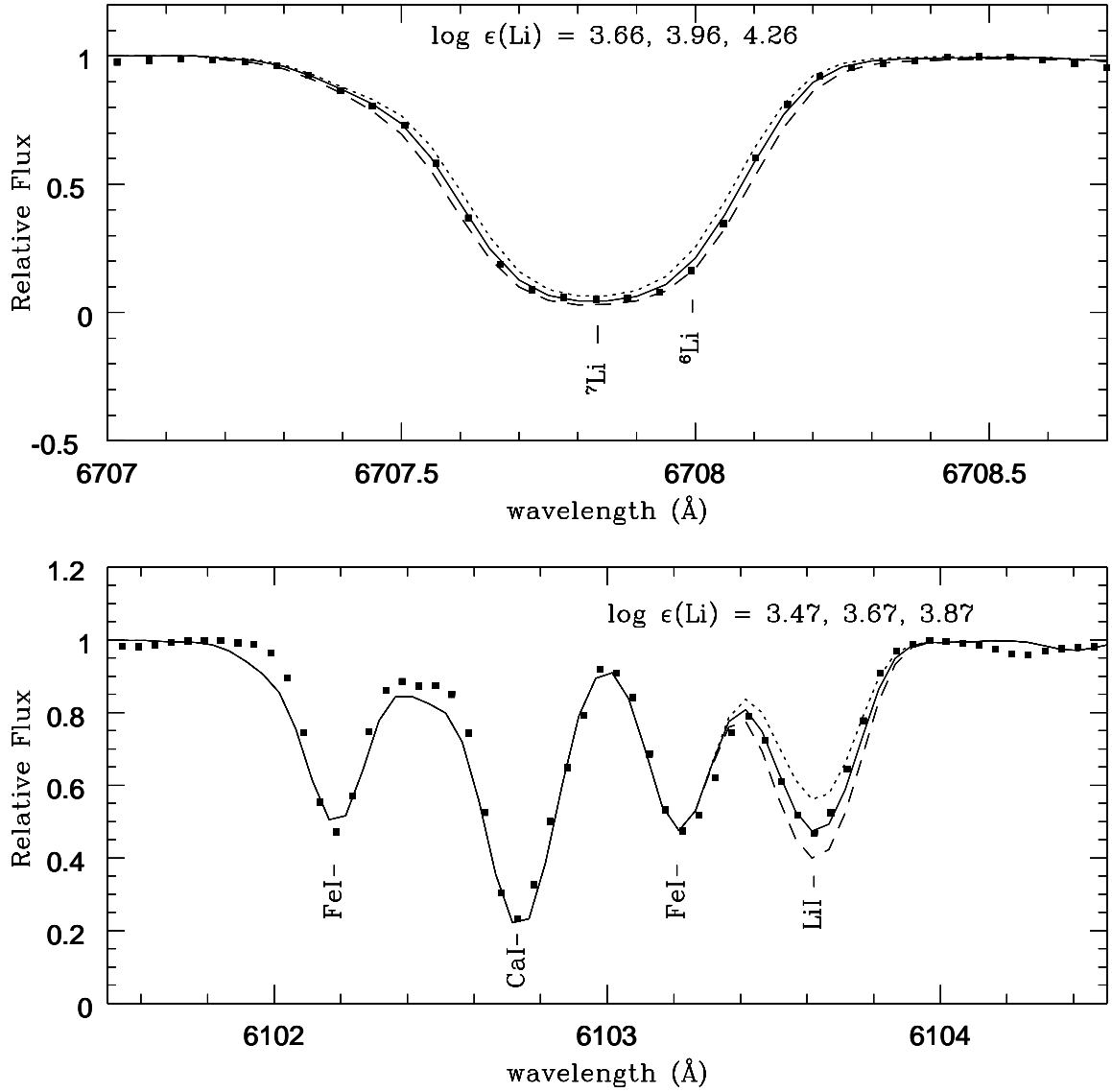}
\caption{Comparison of observed spectrum with the spectrum computed using the LTE model
atmospheres in the wavelength region of Li resonance line at 6707.8\AA\ and excited line at 6103.6\AA.
\label{fig1}}
\end{figure}

\newpage

\begin{figure}
\plotone{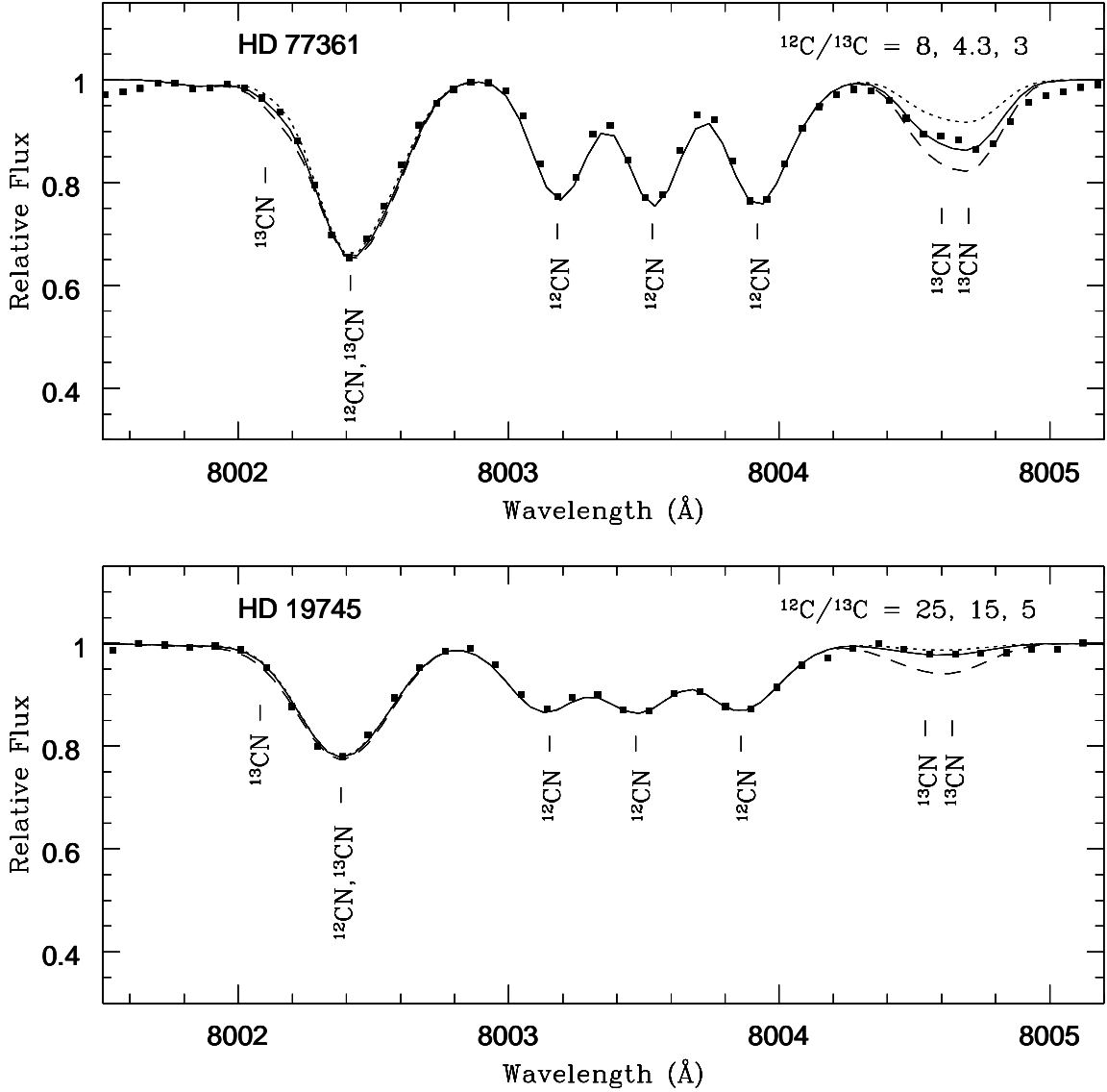}
\caption{Comparison of synthetic and observed spectra for HD~77361 and the template
star HD~19745 in the wavelength region of 8000\AA. Ratio of $^{12}$C/$^{13}$C is derived using 
the $^{13}$C$^{14}$N lines at 8004.6\AA.
\label{fig2}}
\end{figure}

\newpage

\begin{figure}
\plotone{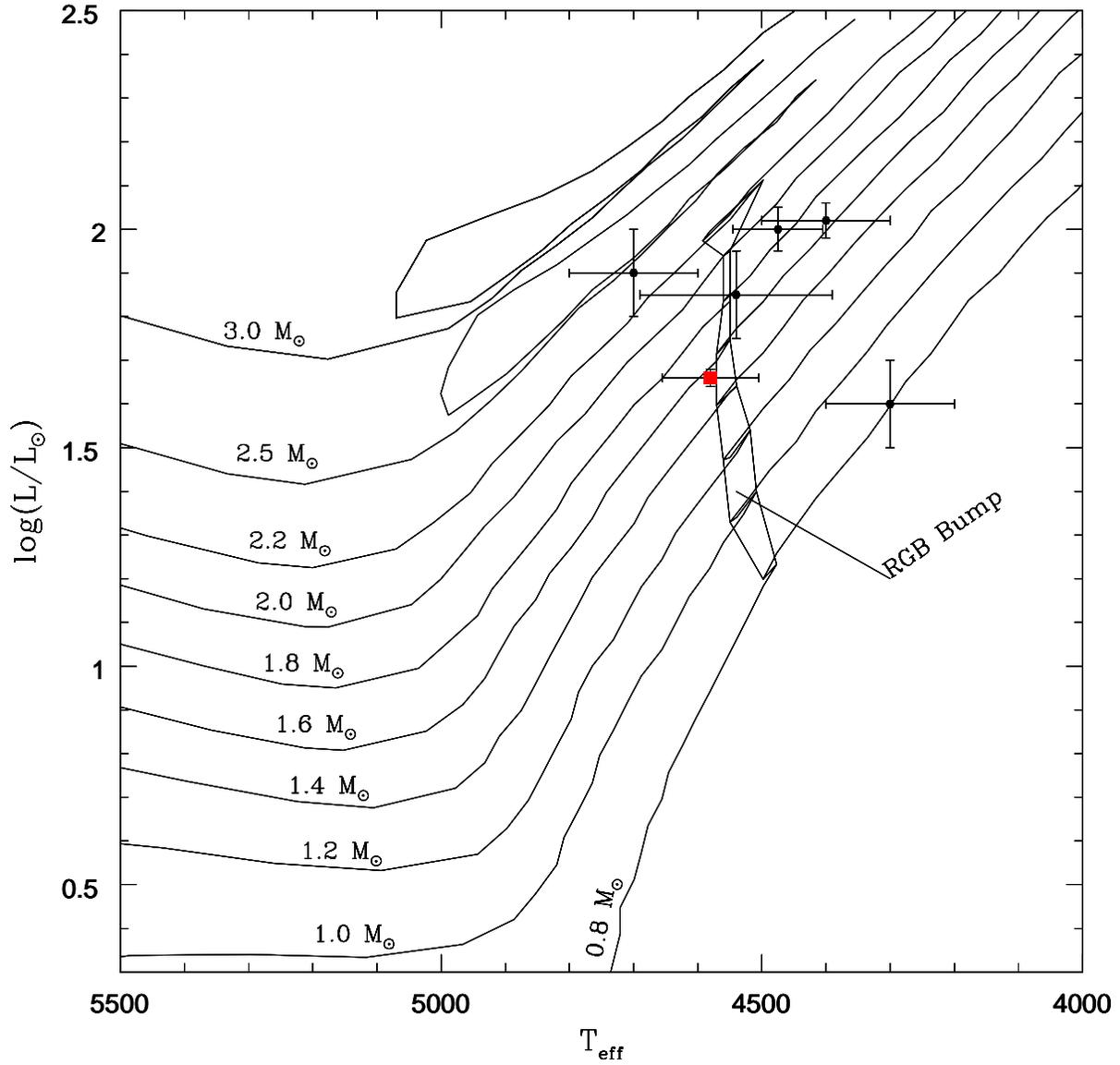}
\caption{Location of super Li-rich K giants on the HR-diagram. Evolutionary tracks of
solar metallicity of masses 0.8 - 3M$\odot$ are plotted. HD~77361 (red square)
and the RGB luminosity bump region are marked.
\label{fig3}}
\end{figure}


\begin{thebibliography}{}
\bibitem[Alexander(1967)]{alexander1967} Alexander J.B., 1967, The Observatory 87, 238
\bibitem[Allende Prieto et al.(2002)]{allende2002} Allende Prieto, C., Asplund, M., {L{\'o}pez}, R.~J.~G., \& Lambert, D. L. 2002, \apj, 567, 544
\bibitem[Alonso et al.(1999)]{alonso1999} Alonso, A., Arribas, S., \& {Mart{\'{\i}}nez-Roger}, C.  1999, \aaps, 140, 261
\bibitem[Balachandran et al.(2000)]{balachandran2000} Balachandran, S. C., Fekel, F. C., Henry, G. W., \& Uitenbroek, H. 2000, \apj, 542, 978
\bibitem[Boothroyd \& Sackmann(1999)]{boothroyd1999} Boothroyd, A.~I., \& Sackmann, I.-J. 1999, \apj, 510, 232
\bibitem[Brown et al.(1989)]{brown1989} Brown, J. A., Sneden, C., Lambert, D. L., \& Dutchover, E. J. 1989, \apjs, 71, 293
\bibitem[Cameron \& Fowler(1971)]{cameron1971} Cameron, A. G. W., \& Fowler, W. A. 1971, \apj,164, 111 
\bibitem[Carlsson et al.(1994)]{carlsson1994} Carlsson, M., Rutten, R. J., Bruls, J. H. M. J., \& Shchukina, N. G. 1994, \aap, 288, 860
\bibitem[Charbonnel(1995)]{charbonnel1995} Charbonnel, C. 1995, \apj, 453, L41+
\bibitem[Charbonnel and Balachandran(2000) ]{charbonnel2000} Charbonnel, C., \& Balachandran, S. C. 2000, \aap, 359, 563
\bibitem[Charbonnel et al.(1998)]{charbonnel1998} Charbonnel, C., Brown, J. A., \& Wallerstein, G. 1998, \aap, 332, 204
\bibitem[de La Reza et al.(1995)]{delareza1995} de La Reza, R., \& da Silva, L. 1995, \apj, 439, 917
\bibitem[de La Reza et al.(1997)]{delareza1997} de La Reza, R., Drake, N. A., da Silva, L., Torres, C. A. O., \& Martin, E. L. 1997, \apj, 482, L77+
\bibitem[Denissenkov \& Herwig(2004)]{denissenkov2004} Denissenkov, P. A., \& Herwig, F. 2004, \apj, 612, 1081
\bibitem[Denissenkov \& Weiss(2000)]{denissenkov2000} Denissenkov, P. A., \& Weiss, A. 2000, \aap, 358, L49
\bibitem[Drake et al.(2002)]{drake2002} Drake, N. A., de la Reza, R., da Silva, L., \& Lambert, D. L. 2002, \aj, 123, 2703
\bibitem[Eggleton et al.(2008)]{eggleton2008} Eggleton, P. P., Dearborn, D. S. P., \& Lattanzio, J. C. 2008, \apj, 677, 581
\bibitem[Girardi et al.(2000)]{girardi2000} Girardi, L., Bressan, A., Bertelli, G., \& Chiosi, C. 2000, \aaps, 141, 371
\bibitem[Gratton et al.(1989)]{gratton1989} Gratton, R. G., \& DâAntona, F. 1989, \aap, 215, 66
\bibitem[Gratton et al.(2000)]{gratton2000} Gratton, R. G., Sneden, C., Carretta, E., \& Bragaglia, A. 2000, \aap, 354, 169
\bibitem[Hobbs et al.(1999)]{hobbs1999} Hobbs, L. M., Thorburn, J. A., \& Rebull, L. M. 1999, \apj, 523, 797
\bibitem[Houk(1982)]{houk1982} Houk, N. 1982, Michigan Catalog of Two Dimensional Spectral Types for the HD Stars, Vol. 3 (Ann Arbor: Univ. Michigan)
\bibitem[Iben(1967a)]{iben1967a} Iben, I. J. 1967a, \apj, 147, 650
\bibitem[Iben(1967b)]{iben1967b} Iben, I. J. 1967b, \apj, 147, 624
\bibitem[Kraft et al.(1999)]{kraft1999} Kraft, R. P., Peterson, R. C., Guhathakurta, P., Sneden, C., Fulbright, J. P., \& Langer, G. E. 1999, \apjl, 518, 53
\bibitem[Kurucz(1994)]{kurucz1994} Kurucz, R. L. 1994, www.kurucz.harvard.edu
\bibitem[Lambert et al.(1980)]{lambert1980} Lambert, D.L., Dominy, J.F., \& Sivertsen, S. 1980, \apj, 235, 114
\bibitem[Lambert \& Reddy(2004)]{lambert2004} Lambert, D. L., \& Reddy, B. E. 2004, \mnras, 349, 757
\bibitem[Lambert \& Sawyer(1984)]{lambert1984}Lambert, D. L., \& Sawyer, S. R. 1984, ApJ, 283, 192
\bibitem[Lind et al.(2009)]{lind2009} Lind, K., Asplund, M., \& Barklem, P. S. 2009, ArXiv e-prints
\bibitem[Lodders(2003)]{lodders2003} Lodders, K. 2003, \apj, 591, 1220
\bibitem[Mallik(1999)]{mallik1999} Mallik, S. V. 1999, \aap, 352, 495
\bibitem[Monaco \& Bonifacio (2008)]{monaco2008}Monaco L., \& Bonifacio P. 2008, Memorie della Societa Astronomica Italiana, 79, 524
\bibitem[Palacios et al.(2001)]{palacios2001} Palacios, A., Charbonnel, C., \& Forestini, M. 2001, \aap, 375, L9
\bibitem[Perryman et al.(1997)]{perryman1997} Perryman, M. A. C., Lindegren, L., Kovalevsky, J., et al. 1997, \aap, 323, L49
\bibitem[Rao et al.(2005)]{rao2005} Rao, N. K., Sriram, S., Jayakumar, K., \& Gabriel, F. 2005, JApA, 26, 331
\bibitem[Reddy \& Lambert(2005)]{reddy2005} Reddy, B. E., \& Lambert, D. L. 2005, \aj, 129, 2831
\bibitem[Reddy et al.(2002)]{reddy2002} Reddy, B. E., Lambert, D. L., Hrivnak, B. J., \& Bakker, E. J. 2002, \aj, 123, 1993
\bibitem[Reddy et al.(2003)]{reddy2003} Reddy, B. E., Tomkin, J., Lambert, D. L., \& Allende Prieto, C. 2003, \mnras, 340, 304
\bibitem[Sackmann \& Boothroyd(1999)]{sackmann1999} Sackmann, I.-J., \& Boothroyd, A. I. 1999, \apj, 510, 217
\bibitem[Siess \& Livio(1999)]{siess1999} Siess, L., \& Livio, M. 1999, \mnras, 308, 1133
\bibitem[Smith \& Lambert (1989)]{smith1989} Smith, V.V., \& Lambert, D.L. 1989, \apjl, 345, 75
\bibitem[Sneden(1973)]{sneden1973} Sneden, C. A. 1973, PhD thesis, AA(The University Of Texas at Austin.)
\bibitem[Sweigart \& Mengel(1979)]{sweigart1979} Sweigart, A. V., \& Mengel, J. G. 1979, \apj, 229, 624
\bibitem[van Leeuwen et al.(2007)]{vanleeuwen2007} van Leeuwen, F., ed. 2007, Astrophysics and Space Science Library, Vol. 350, Hipparcos, the New Reduction of the Raw Data

\end{thebibliography}
\end{document}